%% file: main.tex
\documentclass[conference]{IEEEtran}
\IEEEoverridecommandlockouts
\usepackage{cite}
\usepackage{amsmath,amssymb,amsfonts}
\usepackage{algorithmic}
\usepackage{graphicx}
\usepackage{textcomp}
\usepackage{xcolor}
\usepackage{cleveref}
\usepackage{xurl}
\def\BibTeX{{\rm B\kern-.05em{\sc i\kern-.025em b}\kern-.08em
    T\kern-.1667em\lower.7ex\hbox{E}\kern-.125emX}}

\newcommand{\method}[0]{semantic atlas}

\newcommand{\para}[1]{\vspace{.5mm}\noindent\textbf{#1.}}

\newcounter{JulianaNOC}
\stepcounter{JulianaNOC}

\newcounter{ChristosNOC}
\stepcounter{ChristosNOC}

\newcounter{LauraNOC}
\stepcounter{LauraNOC}

\newcounter{YurongNOC}
\stepcounter{YurongNOC}

\begin{document}

\title{Data Agents Need a Map: Evolving, Demand-Driven Knowledge of Large Collections}

\author{\IEEEauthorblockN{Christos Koutras}
\IEEEauthorblockA{
\textit{New York University}\\
New York, NY, USA \\
christos.koutras@nyu.edu}
\and
\IEEEauthorblockN{Yurong Liu}
\IEEEauthorblockA{
\textit{New York University}\\
New York, NY, USA \\
yurong.liu@nyu.edu}
\and
\IEEEauthorblockN{Laura Haas}
\IEEEauthorblockA{
\textit{University of Massachusetts Amherst}\\
Amherst, MA, USA \\
lhaas@umass.edu}
\and
\IEEEauthorblockN{Juliana Freire}
\IEEEauthorblockA{
\textit{New York University}\\
New York, NY, USA \\
juliana.freire@nyu.edu}
}

\maketitle

\begin{abstract}
Most data that could power rich analysis sits unintegrated and uncataloged in open data portals, data lakes, and collections of independently published tables. Data agents now answer analytical questions over such collections with a full workflow per question, including dataset retrieval, join discovery and evaluation, and tailored program synthesis, most of which are discarded when a new task arrives. The knowledge these workflows depend on, such as what each column means, which context a dataset leaves implicit, and how datasets combine, is therefore perpetually rederived and never persisted.

We argue for persisting this knowledge in an evolving \emph{\method{}} that users and agents can navigate and extend. The atlas describes every dataset offline, through topics, semantic types, and profiles, so that agents can find the datasets that a task requires. Relationships between datasets and integrated views, which require comparing datasets, are instead built on demand for the areas of analysis that users declare. Every item is stored with its provenance and the confidence of the methods that produced it, so that agents can decide how far to trust it or whether to abstain. As datasets change, inconsistencies are detected and corrected automatically, while human users settle what the data leaves inconclusive. We present this vision and its challenges, namely building the atlas over large, multi-domain collections, keeping it correct and up to date, serving it to users and agents, and evaluating whether it improves their answers. A proof of concept on DAB, a benchmark for data agents, shows that with only part of the atlas a coding agent can already read a fifth to a third fewer tokens while achieving the same accuracy as without it.
\end{abstract}


\input{1_Introduction}
\input{2_Building_Atlas}
\input{3_System}
\input{4_Evaluation}
\input{5_Related_Work}
\input{6_Conclusion}

\bibliographystyle{IEEEtran}
\bibliography{references}

\end{document}

%% file: 1_Introduction.tex
\section{Introduction}



Recent advances in large language models (LLMs) have led to systems that aim to answer analytical questions over large collections of independently published datasets, going beyond dataset discovery \cite{fernandez2018aurum, nargesian2018table, zhu2019josie, castelo2021auctus, esmailoghli2025blend}. LLM-based methods now retrieve tables for natural language questions \cite{balaka2025pneuma} and translate questions into SQL over a given database \cite{pourreza2023din, gao2024text}, and \emph{data agents} combine both steps to select datasets, identify how these join or union, and synthesize a program that answers a question \cite{balaka2026pneuma, wang2026ecotable, zhang2026agentic, eckmann2026vision}.

Answering questions over such collections remains challenging. In KramaBench, a benchmark for data science tasks in data lakes, the best system reaches 55\% end-to-end accuracy in a whole data lake, and 62\% when given the relevant files \cite{lai2026kramabench}. Finding the right datasets is only part of the problem, as over 60\% of the failures of a strong coding agent arise from semantic misinterpretation, such as identifier mismatches or invalid proxies for what a question asks \cite{patel2027model}. Analytical questions over such collections require knowledge that no single dataset states, namely i)~what each column means and which columns of different datasets identify the same entities, ii)~the context that a dataset leaves implicit, such as the time period that it covers, which may be stated only in its title or in a publication that describes the data collection process, and iii)~which datasets combine, on which columns, and under what notion of equality. For instance, datasets on NYC Open Data record the six-character code that identifies each school, termed DBN, under many different column names. 

Most data agents derive this knowledge within each task: they retrieve candidate datasets, determine how these combine, i.e., by joins or unions, either through a language model that reasons over the schemas and sample values that it is given or through an external join discovery tool, and synthesize and execute a program over them \cite{balaka2026pneuma, wang2026ecotable, zhang2026agentic}. The knowledge produced along the way, such as which columns to join and under which normalization of their values, or which datasets hold the same data for different years, exists only as an intermediate state of the task, and its correctness is established only by successful execution and inspection of the results. Later tasks that need it must therefore derive it again, although recent systems keep parts of it across queries \cite{wang2026ecotable, balaka2026pneuma, liang2026datareadme}, as we discuss in \Cref{sec:related}. In practice, data is often integrated by hand, so the same work is repeated for each analysis at greater cost; for instance, a single published investigation of NYC school data required weeks of matching school records across publishers in one-off scripts \cite{thecity}, whose results are rarely reusable even by their authors.

A natural response to this cost is to expect that, as agents improve, they will simply absorb this work — building whatever semantic knowledge a task needs from scratch, every time, the way a sufficiently capable end-to-end system might eventually render purpose-built components unnecessary. This view echoes the bitter lesson's appeal to scale over structure, and recent work has asked whether data agents should expect the same trajectory~\cite{patel2027model} Our own experience building dataset discovery, semantic type annotation, schema matching, and harmonization systems suggests otherwise: treating these as problems for a general-purpose agent to solve from scratch is often expensive and unreliable, whereas decomposing them into well-defined components — each combining algorithmic methods with LLM reasoning in ways tailored to what the component must guarantee — has repeatedly proven more effective and more robust~\cite{feuer2024archetype,koutras2026stratyper,zhang2025autoddg,koutras2025omnimatch,liu2025magneto,wu2025bdiviz,freire-vldb2026}. This mirrors a lesson we have learned from software and systems design: durable systems are built from modular, independently engineered and independently optimizable components, not bespoke, monolithic solutions rebuilt for every workload. Data systems should follow the same principle: they should be decomposed into components that can each be built, evaluated, and improved independently, and mixed and matched differently depending on the collection and the task.

Persisting knowledge about data for software that acts on behalf of users is a long-standing goal, from mediators \cite{wiederhold1992mediators} and the Semantic Web \cite{hendler2001agents} to the semantic layers that analysts now author over warehouses and expose to AI agents \cite{dbtsemantic, snowflakesemantic, dbtmcp}. Authoring them does not keep pace with the data; for example, Snowflake reports semantic views covering under 5\% of nearly ten thousand of its own tables even with automated generation \cite{cortexsense}.  Directions such as ontology-based data access and enterprise knowledge graphs (\Cref{sec:related}) assume that people author the semantics, at least for the data that matters, and most assume the data is already integrated, whereas neither holds for independently published collections.

We argue that data agents need what we term a \emph{\method}, i.e., a map of the collection that persists semantic knowledge across tasks and that is built from the data and extended as tasks demand it. Recent work has also argued that, as models improve, the enduring opportunity for data systems lies in \emph{persistent semantic context}, i.e., knowledge about the data environment constructed offline and serving agents across many queries \cite{patel2027model}. However, such context is derived from schemas, documentation, and past queries, which collections of independently published datasets rarely provide. The \emph{dataspaces} line of work deferred creating such knowledge, improving semantic integration as usage and feedback accumulate \cite{franklin2005dataspaces, halevy2006principles, jeffery2008pay}, which was the right response when deciding what a column means and what a dataset records was a semantic judgment that no method could supply at scale. Language models can now supply this judgment, as they assign semantic types across large collections with little supervision \cite{feuer2024archetype, koutras2026stratyper} and describe what each dataset records \cite{zhang2025autoddg}, but they are not reliable; their suggestions can be wrong, and different annotators can assign different types to the same column.

\begin{figure}[t]
    \centering
    \includegraphics[width=\columnwidth]{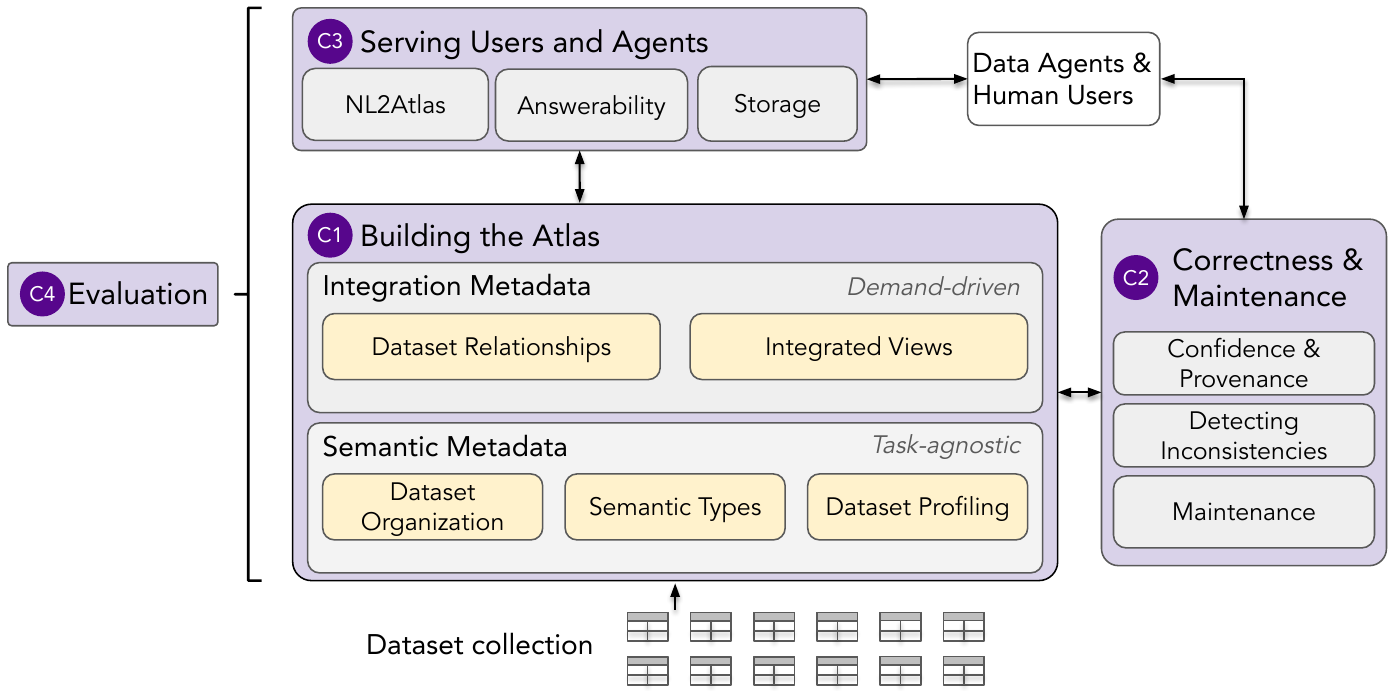}
    \vspace{-.3cm}
    \caption{The \method{} holds task-agnostic semantic metadata and demand-driven integration metadata, and raises four challenges, namely building it (C1), keeping it correct and up to date (C2), serving users and agents (C3), and evaluating it (C4).}
    \vspace{-.4cm}
    \label{fig:atlas}
\end{figure}

\begin{figure*}[t]
  \centering
  \includegraphics[width=\textwidth]{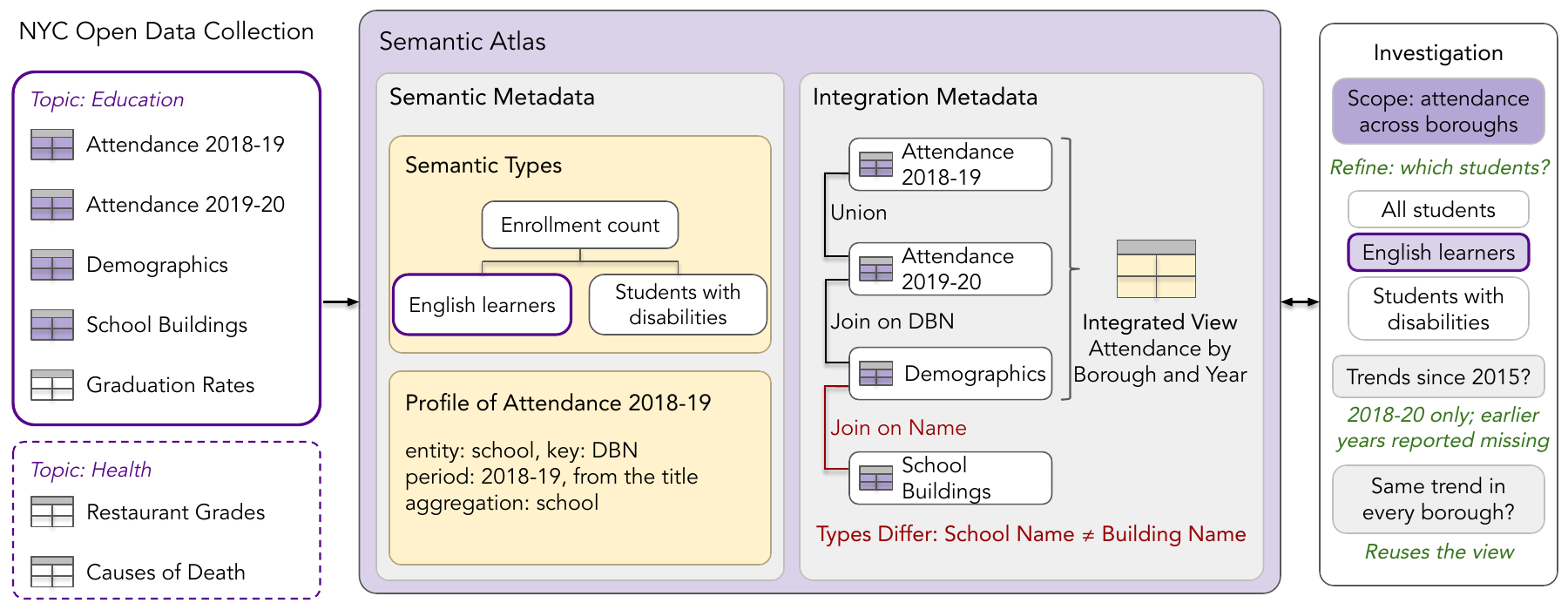}
  \caption{An illustrative example of the \method{} over NYC Open Data for an investigation of school attendance across boroughs, refined to English learners. The investigation reaches the Education topic (solid border) and draws on the datasets in purple.}
  \label{fig:example}
\end{figure*}

We expect the atlas to be used in investigations, in which a user or an agent explores a \emph{scope}, i.e., an area of analysis such as school attendance across boroughs in \Cref{fig:example}, through follow-up questions that widen it and refinements that narrow it (\Cref{sec:serving}). Following the \emph{pay-as-you-go} principle \cite{madhavan2007web, das2008bootstrapping}, the atlas is built in two stages, since comparing values across datasets, as join discovery \cite{koutras2025omnimatch}, schema matching \cite{rahm2001survey, koutras2021valentine}, and entity resolution \cite{christophides2020overview} require, grows quadratically with the number of datasets, whereas describing each dataset does not. First, task-agnostic \emph{semantic metadata} is extracted offline, since agents need it to find the datasets that a task requires. It comprises \emph{topics}, i.e., labeled and possibly overlapping clusters of similar datasets, and the semantic types and dataset profiles derived within each topic, since both the meaning of columns and the context that a profile records, such as the period that a dataset covers, vary across domains. Second, demand-driven \emph{integration metadata}, i.e., dataset relationships and integrated views defined by executable mappings \cite{miller2000schema, haas2005clio}, is built for the datasets of a declared scope once it is specific enough, while any question also adds the relationships between the datasets that it combines. Every item is stored with its provenance, i.e., the methods and the evidence from the data that produced it, and the confidence attached to it by these methods, so that inconsistencies between items can be detected, as when two columns share most of their values but carry unrelated types, and agents can cite the items that they use or abstain when these are missing or produced with low confidence.

\para{Challenges} Keeping knowledge across tasks pays off only if it costs less to store than to re-derive and if later tasks can rely on it, which raises four challenges (\Cref{fig:atlas}). First, the cost of building the atlas must follow need (Challenge~1), which motivates separating task-agnostic semantic metadata, reused by every task, from integration metadata, built on demand for the scopes that users declare, and requires methods for both that scale to large, multi-domain collections (\Cref{sec:building}). Second, the atlas must remain correct and up to date as methods are discovered to have erred and datasets change (Challenge~2), which requires every item to be stored with its provenance and the confidence reported by the methods that produced it, so that inconsistencies can be detected and corrected and agents can decide how far to rely on each item (\Cref{sec:correct}). Third, serving users and agents (Challenge~3) requires i)~mapping each natural language task to the items of the atlas that it needs, which we term NL2Atlas by analogy with NL2SQL, ii)~determining whether these items suffice to answer the task, so that agents can abstain or propose an answerable refinement, and iii)~storing the atlas in forms that both human users and agents can exploit (\Cref{sec:serving}). Finally, evaluating the atlas (Challenge~4) requires measuring whether answers become more accurate, more consistent, and cheaper than when agents build such knowledge themselves, and whether agents abstain on unanswerable questions, whereas benchmarks such as KramaBench \cite{lai2026kramabench} and DAB \cite{ma2026can} pose only independent questions that all have answers (\Cref{sec:evaluation}).

\para{Contributions} This paper makes the following contributions:
\begin{itemize}
    \item We introduce the \method{}, which holds task-agnostic semantic metadata and integration metadata built on demand for declared scopes, and identify, for each of its components, the methods that can build it and the challenges that remain open (\Cref{sec:building}).
    \item We set out the challenges of keeping the \method{} correct and up to date, of serving it to users and agents, and outline directions for each (\Cref{sec:system}).
    \item We discuss different directions for evaluating the \method{} and present a proof of concept on DAB \cite{ma2026can}, in which semantic types, profiles, and relationships let a coding agent read a fifth to a third fewer tokens while answering as accurately as without the atlas (\Cref{sec:evaluation}).
\end{itemize}

%% file: 2_Building_Atlas.tex
\section{Building the Atlas}\label{sec:building}

In this section, we explore how to build the atlas. We
state what each component of the atlas should hold, what existing
methods (if any) can serve as a starting point for building that
component, and where they need to be extended to reach the full vision. 
Across every component below, both algorithm-only and LLM-only approaches fall short on their own, which is itself evidence for treating these as separable, independently improvable parts of a system rather than as a single task to hand an agent.
\Cref{fig:example} illustrates an atlas over NYC Open Data, to which the examples of this section and the next refer.

\subsection{Semantic Metadata}\label{sec:semantic}

\para{Dataset Organization} Every task starts by finding the datasets that it needs, and agents answer less accurately when they must find these datasets themselves within a large collection than when they are given them, as KramaBench shows by comparing the two settings \cite{lai2026kramabench}. Narrowing each task to the few topics that may be relevant to it should therefore improve accuracy, since agents then search only the datasets of these topics rather than the whole collection. A topic is a group of related datasets together with a label and a short summary of what they are about, such as Education and Health in \Cref{fig:example}, so that a task can be matched against the topics rather than against every dataset. Semantic types and dataset profiles should in turn be derived within each topic, since related datasets tend to share types and to differ along the same dimensions of context, such as the school year. As the topics of a collection are rarely known in advance, they must be discovered from the data without comparing every pair of datasets, which is hard for three reasons. First, recall and focus conflict, since a dataset left out of a topic that its data is relevant to is invisible to the tasks that reach this topic, whereas topics must stay small enough to narrow each task and coherent enough to keep the vocabularies of unrelated domains apart. Second, the evidence for grouping is uneven, since descriptions vary in quality across publishers, and titles, schemas, and key column values can suggest different groups. Finally, many datasets belong to several topics, such as a dataset on school meals, which concerns both education and health, so topics must overlap without losing their focus.

Existing methods address parts of this problem. Portal categories, catalogs \cite{halevy2016goods}, and navigation structures for data lakes \cite{nargesian2020organizing} organize datasets for human users who browse them, topic modeling \cite{blei2003latent, grootendorst2022bertopic} can discover topics and the terms that characterize them at low cost from titles, descriptions, and schemas, domain discovery \cite{ota2020data} finds the sets of values that columns share across datasets, such as the DBNs of schools, and language models can label and summarize the groups that result. Agents can inspect far more candidate datasets per task than human users, so structures that favor recall over compactness may serve them better, which remains to be studied. None of these methods, however, addresses all three difficulties, so it remains open how to group datasets into topics, and how to place added datasets in existing or new topics without regrouping the collection, which would require rederiving the types and profiles of every topic.

\para{Semantic Types} Column names are an unreliable guide to what columns mean, since publishers name the same attribute differently and different attributes alike. Semantic types address this by describing the kind of entity or value that a column holds independently of its name, so that, for example, the DBN receives one type under all of its column names. The structure of a semantic type should in turn let a request at any level of generality reach the columns that answer it; for instance, counts of English learners and of students with disabilities are both enrollment counts in NYC Open Data (\Cref{fig:example}). 

Again, existing methods get us part-way to our goal. Most methods for semantic type annotation label the columns of each table without regard to the rest of the collection. They can therefore assign different types to the same attribute across datasets and runs, and the same type to different attributes, such as school names and the names of school buildings, whereas the atlas needs one vocabulary across the datasets of a topic. Supervised annotation methods assign types from fixed vocabularies \cite{zhang2020sato, suhara2022annotating}, which may lack the types of a specific domain, and LLM-based methods such as ArcheType need no training data but require a label set in advance \cite{feuer2024archetype}, without which LLMs produce free-form labels that vary across columns. StraTyper \cite{koutras2026stratyper} discovers the types that a collection records at low cost, but it annotates only some of the columns, so its types must be reconciled with those that LLMs assign to the rest. In addition, annotation methods do not control the granularity of the types that they produce, so counts of English learners and enrollment counts can end up as unrelated types. Organizing the types into a hierarchy would relate them across levels of generality, but it is not straightforward, since a type often falls under several more general types rather than a single parent. Taxonomy induction \cite{zeng2024chain} and ontology learning \cite{giglou2023llms4ol} with LLMs can propose a structure, but they relate terms through their names rather than through the data, so imposing a structure on types remains an open challenge. Finally, the types of different topics must be reconciled, since a scope can draw on datasets from several topics.

\para{Dataset Profiling} Semantic types describe columns, but deciding whether two datasets can be combined also requires knowing what each dataset records as a whole, which agents typically infer from samples in every task. A profile should characterize each dataset, identifying the kind of entity the rows  describe, noting its keys and integrity constraints, computing quality indicators such as the fraction of missing values, and imputing the context that may be implicit, such as the time period covered by the dataset or its level of aggregation. Context is rarely stated as an attribute; for instance, NYC Open Data contains datasets on schools. These datasets have the same columns but for different school years, which only their titles identify, so agents that union two or more of these datasets without adding in the year mix the years, as with the two attendance datasets in \Cref{fig:example}. 

Many existing methods can get us started. Profiling algorithms discover keys, dependencies, and quality indicators \cite{abedjan2015profiling}. AutoDDG generates descriptions of datasets \cite{zhang2025autoddg}, Auctus derives their temporal and spatial extent from the values \cite{castelo2021auctus}, and Data.Readme documents how similar-looking tables differ, for instance in granularity or time period \cite{liang2026datareadme}. However, these methods can stumble on a range of obstacles: dependencies that hold on the current values may be accidental, context is often stated only in titles or descriptions rather than in the values, and the dimensions of context that matter differ across topics. Semantic types can help with the first two obstacles, since they indicate which columns can form meaningful keys and which datasets record the same attributes, so that their titles can be compared to recover the dimension in which they differ. Recovering context within a topic and checking it against the values nevertheless remains an open problem, since the dimensions that matter must be discovered for each topic rather than be fixed in advance. 

\subsection{Integration Metadata}\label{sec:integration}

\para{Dataset Relationships} Many analytical questions over a collection combine several datasets, and agents today identify how these datasets combine in every task, incurring the cost of this step repeatedly. Each discovered relationship should therefore be recorded together with the corresponding columns, the notion of equality and the degree of overlap between their values, and its provenance, so that later tasks can reuse it instead of comparing values again.

Many methods that discover relationships rely on automated matching, whose scores are interpretable only within the run that produced them, whereas a stored relationship must remain meaningful across tasks. In addition, a correspondence that matching returns, i.e., a pair of columns deemed to match, may support a direct join, a join only between rows that share a context, an approximate join that must be reviewed, or only a union, when the columns share a domain but not their values. Join discovery \cite{fernandez2018aurum, zhu2019josie, koutras2025omnimatch}, schema matching \cite{rahm2001survey, koutras2021valentine, liu2025magneto}, and joins beyond exact equality \cite{zhu2017auto, li2021auto} can all propose candidates. Data.Readme indexes joinable and unionable tables after checking the value containment of candidate keys and the plausibility of joined samples \cite{liang2026datareadme}, although it leaves the disambiguation of plausible relationships to each question. Learned matchers such as OmniMatch~\cite{koutras2025omnimatch} are effective mainly within the domains of their benchmarks, and although LLM-based matchers generalize across domains, both decide correspondences without knowing what the datasets record, so columns that look alike but come from unrelated domains yield false matches. 

In our vision, the semantic types and profiles of the topics in a scope can focus matching on plausible pairs, and a disagreement between types and values can reveal a false match or a wrong type, as for the join on names in \Cref{fig:example} (\Cref{sec:correct}). However, deciding whether a correspondence permits only a union or a join, and under which conditions, remains open.

\para{Integrated Views} Agents often answer the questions of a scope from the same datasets, yet they currently combine these datasets with a tailored program for each question. An integrated view, such as attendance by borough and year in \Cref{fig:example}, should instead assemble the datasets of a declared scope once, through an executable mapping in the sense of Clio \cite{miller2000schema, haas2005clio}, stored with its provenance, so that later tasks in the scope query the view rather than assemble the data again. Determining the correct view is hard, since several combinations of datasets can serve a scope with different completeness and cost, corresponding values may differ in format or meaning, the context of each dataset decides whether datasets must be joined or unioned, and datasets often disagree on the values that they report for the same entities. Existing methods address parts of this problem: Clio itself derives mappings from correspondences that humans provide, while methods for integrating data lake tables combine tables automatically, at opposite ends of the trade-off between completeness and cost. ALITE integrates a given set of tables through full disjunction, which keeps every tuple and combines those that agree on shared columns \cite{khatiwada2022integrating}, whereas EcoTable integrates only the tables that a natural language query needs, at low cost, but considers only joins \cite{wang2026ecotable}. Data fusion then resolves the conflicts that remain when datasets report different values for the same entities \cite{bleiholder2009data}. Deciding when values agree, however, and whether to resolve a conflict, carry it into the view, or decline the view, often requires knowledge that the collection lacks, so human users must take part in these decisions (\Cref{sec:correct}). Together, these methods can assemble preliminary views, but as they integrate a given set of tables or serve a given query, they leave open which views are worth keeping for later tasks in a scope. Since each kept view must be profiled and typed like any other dataset, so that tasks can find it, and kept consistent with the datasets that it combines, a view is worth keeping only if enough later tasks use it.

%% file: 3_System.tex
\section{Keeping the Atlas Correct and Serving It}\label{sec:system}

In this section, we explore how to keep the atlas correct and up to date and how to serve it to human users and agents. Whereas the components of \Cref{sec:building} produce the items of the atlas, those of this section operate on them, through programmatic methods, agents, or both. For each component, we state its goal, the existing work that it can build on, and the problems that remain open.

\subsection{Correctness and Maintenance}\label{sec:correct}

\para{Confidence and Provenance} Human users and agents can trust an item that they did not build only if they know how it was produced and how reliable it is; otherwise, agents derive the item again in every task, and human users may accept it unchecked. Every item should therefore be stored with its provenance, i.e., the datasets and items from which it was built and the methods and evidence that produced it, and with the confidence that these methods place in it, so that both can decide how far to rely on it. For instance, the provenance of a relationship might include the containment of the DBNs of one dataset in those of the other, which an agent can recheck if it doubts the relationship. Confidence, however, is hard to quantify, let alone on a common scale, since methods report it in incomparable forms, if at all, and some items, such as the equivalence of two types, admit no direct check against the values. In addition, the confidence that a task requires depends on the consequences of an error, which grow with the number of tasks that reuse an item.

Data provenance explains why and where the results of a query come from \cite{buneman2001why}, but it concerns deterministic queries, whereas the methods that build the atlas can produce false results. Selective prediction \cite{geifman2017selective} and conformal prediction \cite{angelopoulos2023conformal} decide when a model should abstain, and semantic operators guarantee accuracy with respect to a reference algorithm \cite{patel2025semantic}, but they assume labeled data or a reference, which a new collection lacks. Items that the data leaves inconclusive should go to human users, for whom feedback requests can be ordered by their expected value \cite{jeffery2008pay} and interfaces such as BDIViz present the evidence behind each candidate match \cite{wu2025bdiviz}. A key challenge is thus to quantify confidence, even where methods report none, on a scale common to all methods, and another is to present provenance so that agents can recheck it while human users judge it without simply deferring to the methods.

\para{Detecting Inconsistencies} Items that different methods build can each seem plausible yet contradict each other, and such a contradiction can reveal an error even when each method is confident in its own item. Programmatic checks should therefore flag conflicts between items systematically, such as two columns that share most of their values but carry unrelated types, or several pairs of columns with the same type whose values never overlap. A flagged conflict, however, does not show which item is wrong, or whether any is; for instance, unrelated types on overlapping columns can reveal a false match or a wrong type, whereas columns of the same type without common values may simply hold disjoint subsets of a domain, such as the DBNs of schools in different boroughs. Agents must reason over the provenance of the conflicting items, and over how confident their methods were, to propose a correction, while human users settle the conflicts that the data leaves inconclusive. 

Automated methods in the literature resolve similar conflicts in related settings. Schema matching networks detect correspondences that violate consistency constraints across many schemas and prioritize the least certain ones for validation by human users \cite{nguyen2014pay}, and HoloClean repairs values that violate integrity constraints through probabilistic inference \cite{rekatsinas2017holoclean}. These methods, however, rely on constraints that are given in advance, whereas the conflicts in the atlas arise between items of different kinds, such as types and relationships, for which few constraints are given. This leaves open which conflicts to check and how much evidence a correction requires when many tasks reuse the corrected item.

\para{Maintenance} Collections change as datasets are added, revised, and removed, so items built for earlier tasks may no longer hold when later tasks reuse them. Each change should therefore update the items that it affects, i.e., the items that depend on the changed dataset or item. Finding them requires a \emph{dependency graph} that links each dataset and item to the items that depend on it, much as the system catalogs of relational databases record which views depend on each table. For instance, a revision of an enrollment dataset could lead through this graph to its relationships, whose associated value overlap may change, and to the integrated views built on them. 

Dependencies in the atlas, however, are less simple than those in database catalogs, where the definition of a view names the tables on which it depends. Besides the datasets and items from which it was built, an item can also depend on other ones that guided its construction without being among its inputs; for instance, a relationship depends on the two columns that it relates, but also on the types that focused matching on these columns and on the profiles that determined whether they join or only union. Changes can be detected programmatically, by comparing each new version of a dataset with its recorded profile, while an agent that monitors the collection can decide which affected items to recheck first.  

Maintaining the atlas can build on the principles of seminal work on propagating change, such as the retraction of conclusions whose justifications no longer hold in truth maintenance systems \cite{doyle1979truth}, the incremental update of materialized views as their base data change \cite{gupta1995maintenance}, and the adaptation of mappings \cite{velegrakis2003mapping} and ontologies \cite{noy2004ontology} as schemas and domains change. These principles, however, were developed for exact dependencies and declared changes, whereas the dependencies of the atlas may be incomplete, its items rest on evidence of varying strength, and datasets are often revised without a description of what changed. Future work must determine which dependencies to record for each item, and when a change should lower the confidence in an item rather than remove it.

\subsection{Serving Users and Agents}\label{sec:serving}

\para{NL2Atlas} Tasks arrive in natural language, whereas the atlas holds topics, types, profiles, relationships, and views, so NL2Atlas should map each task to the items that it needs. For instance, a task on school attendance across boroughs should reach the attendance datasets and the columns from which the borough of each school follows, such as the DBN, which encodes it. The wording of a task follows the vocabulary and level of generality of its author rather than those of the collection, so its terms must be matched to types at the right level of the type hierarchy. Users also often declare scopes that are too broad to integrate affordably, such as education in NYC Open Data, and agents should then refine such a scope with them in successive steps, along the topics, types, time periods, and levels of aggregation that the atlas records, for instance from education to school attendance across boroughs and then to a range of school years. Human users can weigh only a few refinements per step, whereas agents can pursue several in parallel, so the options offered to human users should minimize their effort, as faceted search over structured databases does \cite{basu2008minimum}. 

Both steps of NL2Atlas, mapping and refinement, have counterparts in narrower settings. For the mapping, dataset discovery finds the datasets relevant to a query \cite{fernandez2018aurum, nargesian2018table, castelo2021auctus} and NL2SQL translates a question into a query over a known schema \cite{pourreza2023din, gao2024text}, while for the refinement, clarifying questions narrow ambiguous requests in conversational search \cite{aliannejadi2019asking}, data exploration helps users who do not yet know what they are looking for \cite{idreos2015overview}, DataScout reformulates dataset searches based on the datasets that they return \cite{lin2025rethinking}, and Ver lets users choose among candidate views \cite{gong2023ver}. These methods, however, map or refine a single request, whereas NL2Atlas must map tasks across many partially integrated schemas and support the refinement of scopes that persist across questions. The open problems include matching a task at the right level of generality and choosing the refinements to offer at each step.

\para{Answerability} Agents produce plausible answers even when the data cannot support them, so an agent should first determine whether the items that NL2Atlas returns answer a task fully, partially, or not at all. A fully answerable task then costs a lookup, a partially answerable one requires building only its missing items, and for an unanswerable one the agent should abstain or propose the nearest answerable refinement, citing in every case the items that it uses. For instance, an agent asked for attendance trends since 2015 should state for which school years the collection holds attendance data and offer the trend over these years, rather than extrapolate beyond them.

Deciding full answerability resembles answering queries using views \cite{halevy2001answering}, which determines whether a query can be answered from stored views alone, i.e., whether these views contain all that the query needs. The same containment reasoning applies across scopes, since the questions of a narrower scope can be answered from the relationships and views built for its parent; for instance, attendance in a single borough can be read from the view of attendance by borough and year, whereas widening the scope to graduation rates calls for new relationships and views. Selective question answering abstains when a model is likely to be wrong \cite{kamath2020selective}, text-to-SQL methods detect questions that a schema cannot answer \cite{wang2023know}, conversational benchmarks pair such questions with helpful responses \cite{dong2025practiq}, and query refinement adjusts queries that return too few answers \cite{mishra2009interactive}. These methods decide from the confidence of a model, a fixed schema, or a single query, whereas answerability over the atlas is a matter of degree, since an answer may rest on items produced with varying confidence and built for only part of the collection. Two problems stand out, namely estimating how far the items of the atlas support an answer, and reporting what is missing so that refinement does not reduce the analysis to what the data happens to support.

\para{Storage} The atlas helps agents only if they can read the items that a task needs, and agents increasingly read context as Markdown files loaded into their prompt, such as AGENTS.md files \cite{agentsmd} and agent skills \cite{li2026skillsbench}, which they can also curate for themselves as they work \cite{zhang2026ace}. Files, however, suit context that fits in a prompt, whereas the items of an atlas over a large collection far exceed what an agent can read, and a task needs only those of its topics or scope. One option is to keep the atlas in a dedicated store, i.e., a database that agents query through tools to retrieve only the items that a task needs, such as the types and relationships of one topic. Another is to split the atlas into files, for instance one per topic or scope, that agents open only when a task reaches them. Recent systems already expose context to agents through tools \cite{liang2026datareadme, zhang2026agentic} or through the Model Context Protocol, as semantic layers do \cite{dbtmcp}, but they serve curated or task-specific context, whereas the atlas holds items that methods produced with varying confidence, together with their provenance, across a whole collection. Human users, in contrast, need visual summaries of the evidence behind each item, as BDIViz offers for matches \cite{wu2025bdiviz}, so the same items must be stored once and presented differently to agents and to human users. It remains to be explored whether agents need a dedicated store or files suffice, and how the two could be combined and kept consistent as the atlas changes.

%% file: 4_Evaluation.tex
\section{Evaluating the Atlas}\label{sec:evaluation}

We discuss how to evaluate whether the atlas makes the answers of agents more accurate, more consistent across runs, and cheaper, and whether agents abstain on questions that the collection cannot answer. In what follows, we first set out what such an evaluation requires and why existing benchmarks fall short, and then present a proof of concept on DAB\cite{ma2026can}, in which semantic types, profiles, and relationships guide a coding agent.

\subsection{Requirements}\label{sec:requirements}

The requirements of such an evaluation concern what a benchmark contains, what it measures, and the baselines against which it compares the atlas.

\para{Benchmark Contents} The benefits of the atlas accrue across tasks, since items built for one task serve the later tasks of the same or related scopes, so a benchmark must pose sequences of related tasks rather than independent questions. In the investigation of \Cref{fig:example}, for instance, a user declares a scope on school attendance across boroughs, refines it to English learners, and asks for trends since 2015 before asking whether every borough follows the same trend. Such sequences should also narrow and widen their scope, since a narrower scope can reuse the items built for its parent, whereas a wider one needs items that have yet to be built. The tasks must further include questions that the collection cannot answer, such as questions on school years or attributes that no dataset records, and the collection must come with snapshots taken over time, so that revised and new datasets can be replayed against items built on earlier versions. Existing benchmarks offer none of these, since NL2SQL benchmarks pose questions over known schemas \cite{yu2018spider, li2024can}, and benchmarks for data agents pose independent questions over data lakes or multiple databases and score each one on its own \cite{lai2026kramabench, ma2026can}. Benchmarks with unanswerable questions exist for reading comprehension \cite{rajpurkar2018know} and for text-to-SQL over a known schema \cite{wang2023know, dong2025practiq}, but not for dataset collections, where a question is unanswerable only if no combination of datasets supports it, which is harder to establish than the absence of an answer from a single text or schema. Evaluating the atlas therefore calls for a new benchmark rather than an extension of existing ones.

\para{Metrics} Accuracy and cost should be measured along each sequence, since a question should cost less once the items built for earlier ones are stored, and the cost of building the atlas should be counted as well, so that the savings of later tasks are weighed against it. Since the language models that drive agents can generate different steps for the same input, the same question can receive different answers across runs, so each task should be run several times to measure consistency, for instance as the fraction of runs that return the most frequent answer. Abstention should be scored by relating the error rate of the answers given to the fraction of questions answered \cite{geifman2017selective}, and an answerable refinement offered in place of an abstention should be credited, provided that it stays faithful to what the user asked. The items of the atlas should also be evaluated, since an agent may answer correctly despite wrong items or fail despite correct ones. Types can be evaluated against curated annotations \cite{koutras2026stratyper} and relationships against matching benchmarks \cite{koutras2021valentine}, while confidence should be evaluated by whether methods are more often right when they are more confident, and by the effort that human users spend on the items that the data leaves inconclusive. Curated ground truth is expensive for multi-domain collections, but since each task uses only part of the atlas, it is needed only for the topics and scopes that tasks reach.

\para{Baselines} We consider the components of \Cref{sec:building} necessary for answering questions within a declared scope, but it remains open how much they help agents, and whether agents benefit more from the atlas or from building such metadata themselves, in structures that they handle better. Recent work shows that agents can author such context, but not always to their benefit \cite{patel2027model, ang2026agentic, zhang2026ace, li2026skillsbench}. A benchmark should therefore compare agents that use the atlas with agents that use no stored metadata and with agents that build and keep their own, over the same sequences of tasks, so that the value of each component is measured against what agents could produce without it.


\subsection{Preliminary Results}\label{sec:preliminary}

\para{Setup} As a first step, we build three components of the atlas, namely semantic types, profiles, and relationships, in advance for a whole collection, and test whether they help a coding agent answer questions over the collection. We use DAB \cite{ma2026can}, whose 104 questions span 17 datasets (in DAB, a dataset is a group of databases with related data, e.g., on sales, software vulnerabilities, or stock markets). For each question, DAB connects the agent only to the databases of the question's dataset, so the agent searches a few databases rather than a whole collection. Since the atlas targets large collections, we merge all datasets into one collection of 2,854 tables, so that the agent must also find the databases and tables that a question needs. An LLM assigns semantic types to the columns, reconciles them into 607 types, and places these under 132 parents. Each table receives a profile that combines its documentation with a description generated by AutoDDG \cite{zhang2025autoddg}, its keys, and its coverage. Relationships are derived from the values: for each pair of columns in different tables whose types are equal or share a parent, we measure how many values of each column the other contains. The agent reads the atlas as one Markdown file per table and searches all of its items with a tool. We serve relationships in two ways. On request, the agent lists the relationships of a table with a command, while in place, the file of each table and every search result that returns it also show its joins to keys in its dataset, with how many of its values each key contains and, when some are missing, an example from each side. We use Claude Code \cite{claudecode} with Claude Sonnet 5.5 as the agent, because coding agents explore data by writing programs, and on DAB, a coding agent already outperforms a data agent designed by hand \cite{patel2027model}. We run the agent three times on each question in three settings: without the atlas, with relationships on request, and with relationships in place.

\begin{figure}[t]
  \centering
  \includegraphics[width=\columnwidth]{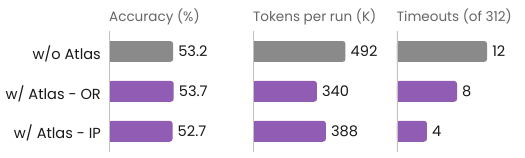}
  \caption{Accuracy, tokens read per run, and runs that exceed the 30-minute limit on DAB, without the atlas and with it, where the agent receives relationships on request (OR) or in place (IP). Part of the atlas already lets the agent answer as accurately as without it while reading 21\% to 31\% fewer tokens.}
  \label{fig:results}
\end{figure}

\para{Results} We measure accuracy with the validators of DAB and cost by the tokens that the agent reads per run. As \Cref{fig:results} shows, the atlas lets the agent answer as accurately as without it at a lower cost: accuracy stays within a point of the 53.2\% that the agent reaches without the atlas, while an average run reads 31\% fewer tokens with relationships on request and 21\% fewer with them in place. Relationships on request cost less than relationships in place; we believe this happens because the agent retrieves them only when it needs a join, whereas in place they accompany every file and search result that returns a table, even when the question needs no join. The actual savings are larger, since runs that exhaust the 30-minute limit report no tokens, and the atlas reduces these runs by a third with relationships on request and by two thirds with them in place. 

\para{Case Studies} 
Without the atlas, the agent must explore the collection before it can answer, and the atlas saves much of this exploration. When asked whether a sales lead qualifies according to its call transcripts, the agent without the atlas reads the documentation of six databases, lists their tables, and probes the database server before it finds the transcripts, spending 83 to 116 thousand tokens per run. With the atlas, a single search returns the column of transcripts under its type, timestamped sales call transcripts, so the agent reads the transcripts of the lead in its next step and spends 28 to 66 thousand tokens, answering correctly in every run (as it does without the atlas). The largest savings occur in the CVE dataset, where the identifiers of vulnerabilities are obfuscated in many ways, so the agent must decode them before it can join two tables. The atlas records that the identifiers of affected products and those of exploited vulnerabilities share no values, and with it, an average run on this dataset reads 2.8 to 2.9 million tokens instead of 5.4 million.

Relationships also prevent errors that the agent would not notice. When support cases are joined to accounts, only 55\% of their identifiers match, because some carry a \texttt{\#} prefix, so the join returns some rows while silently dropping the others. The relationship between the two columns records this mismatch with an example. All three runs that ask for this relationship strip the prefix, compared with one of the three runs without the atlas, contributing to the agent answering correctly. 

\para{Missing Components} This prototype builds part of the atlas, and its other components should save more. Topics would narrow each search to the datasets of a few domains, and confidence would tell the agent how far to trust each item. Integrated views would go further, since the agent decodes the CVE identifiers from scratch for every question that joins them, whereas a view would decode them once for all later questions (\Cref{sec:integration}). Measuring such savings requires sequences of related questions, and measuring abstention requires questions that the collection cannot answer, which is why the benchmark that we call for in \Cref{sec:requirements} would include both.

%% file: 5_Related_Work.tex
\section{Related Work}\label{sec:related}

\para{Data Agents and Persistent Context} Data agents typically build the knowledge that a question needs while answering it, and some keep part of it. For example, EcoTable builds a joinability graph that serves a set of queries \cite{wang2026ecotable}, Pneuma retains distilled notes from earlier interactions \cite{balaka2026pneuma}, and Metadata Reasoner turns data catalogs into context layers that agents query \cite{zhang2026agentic}. Data.Readme indexes joinable, unionable, and confusable tables offline and exposes them to agents through tools, which improves their accuracy on large data lakes \cite{liang2026datareadme}. Recent work argues that persistent semantic context will remain valuable as models improve, while showing that context that agents write for themselves does not always help \cite{patel2027model}.  Visions of agent-first data systems call for a memory store that keeps grounding information for reuse across requests \cite{liu2026supporting} or replace upfront integration with collaborating agents \cite{eckmann2026vision}. We share the goal of these systems and differ in specifying what persistent metadata must hold for large dataset collections, in building its integration metadata only for the scopes and questions that need it, and in recording the provenance of each item and the confidence of the methods behind it, so that inconsistencies can be detected and corrected, which offers one answer to the open problem of keeping such context consistent.

\para{Semantic Layers, Catalogs, and Knowledge Graphs} Mediators introduced a layer of knowledge between applications and independently managed data \cite{wiederhold1992mediators, garcia1997tsimmis}, and the Semantic Web envisioned agents that act on shared ontologies \cite{hendler2001agents}. Industrial semantic layers realize part of this vision over warehouses, where analysts author the entities, relationships, and metrics through which tools and AI agents query integrated tables \cite{dbtsemantic, snowflakesemantic, dbtmcp}, although authoring does not keep pace with the data \cite{cortexsense}, and Auto-Ontology drafts terms and their SQL definitions automatically but still has reviewers certify them \cite{autoontology}. Ontology-based data access depends on hand-written mappings, whose generation is the acknowledged bottleneck \cite{poggi2008linking, xiao2018ontology, calvanese2016ontop}, enterprise knowledge graphs require dedicated teams to maintain them \cite{noy2019industry, hogan2021knowledge}, and knowledge bases such as YAGO are extracted from curated resources \cite{suchanek2007yago}. Data catalogs organize datasets for human users who browse them \cite{halevy2016goods}, Ground proposed a service that collects and serves the context of data, i.e., how it is interpreted, how it was produced and used, and how it changed across versions \cite{hellerstein2017ground}, and schema.org metadata lets agents retrieve usable datasets more precisely, yet each publisher writes it for its own datasets \cite{chen2026agents}. We share the separation of metadata from data with this line of work and differ in deriving the metadata from the datasets of a collection, together with the evidence behind each item, so that human users settle only what the data leaves inconclusive.

\para{Dataspaces and Pay-as-You-Go Integration} Dataspaces offer basic services over a collection immediately and improve integration incrementally as usage and feedback accumulate \cite{franklin2005dataspaces, halevy2006principles}, since integration at web scale can only afford to pay as it goes \cite{madhavan2007web}, with user feedback confirming candidate correspondences \cite{jeffery2008pay} and probabilistic mediated schemas bootstrapping such systems \cite{das2008bootstrapping}. We keep the principle that knowledge accumulates through use and differ in what can precede use, since language models now supply semantic types and dataset descriptions before any task. We also differ in what integration pays for, since integration metadata is built only where scopes or questions call for it and is kept with its provenance. The discovery and integration methods on which the atlas could build are discussed with each of its components in \Cref{sec:building}.

%% file: 6_Conclusion.tex
\section{Conclusion}

In this paper, we argued for a \method{}, a persistent structure that records what the datasets of a collection contain and how they combine, so that human users and agents can navigate and extend this knowledge instead of rederiving it for every task. Topics, together with the semantic types and dataset profiles derived within each topic, are extracted offline, whereas dataset relationships and integrated views are built on demand for the scopes that users declare. Every item is stored with its provenance and the confidence that its methods assign to it, so that users and agents can decide how far to trust it and inconsistencies can be detected, while human users settle only what the data leaves inconclusive. We set out the challenges of building the atlas over large, multi-domain collections, keeping it correct as datasets change, serving it to human users and agents, and evaluating it. A proof of concept on DAB indicates that part of the atlas already lets a coding agent answer as accurately as without it while reading a fifth to a third fewer tokens.